\documentstyle[preprint,aps]{revtex}

\begin{document}

\preprint{DOE/ER/40561-173--INT-94-00-77 and INP-1683/PH}

\title{Pseudo-Goldstone Modes in Isospin-Asymmetric Nuclear Matter}

\author{Thomas D. Cohen\footnote{On leave from Department of
Physics and Astronomy, University of~Maryland, College~Park, MD~20742}}

\address{Department of Physics and Institute for~Nuclear Theory,
University of Washington, \\Seattle, WA~98195, USA}

\author{Wojciech Broniowski}

\address{H. Niewodnicza\'nski Institute of Nuclear Physics,
         PL-31342 Cracow, POLAND}

\maketitle

\begin{abstract}

\noindent
We analyze the chiral limit in dense isospin-asymmetric nuclear matter.
It is shown that the pseudo-Goldstone modes in this system are
qualitatively different from the case of isospin-symmetric matter.

\end{abstract}

\newpage

% begin the body

The concept of mesons modified by the
nuclear medium has been used for many years \cite{mesons:in:medium}.
It has been generally assumed that although the properties of
pions and  kaons (mass, decay constant, coupling constants, etc.)
are changed by the presence
of the medium, their {\em pseudo-Goldstone nature} is
preserved.\footnote{A mode can be identified as a pseudo-Goldstone
mode, if its excitation energy
vanishes in the chiral limit of zero quark mass.} In the vacuum,
the masses of pions or kaons are known to be proportional
to the square root of the current
quark masses, according to the Gell-Mann--Oakes--Renner (GMOR)
relation \cite{GMOR:ref}.
However, in nuclear matter this need not be the case.

In this paper we show, that the nature
of {\em charged} pionic excitations in a dense {\em isospin-asymmetric}
medium, or kaonic excitations
in a non-strange medium, is radically different than what is seen in
the vacuum results.  With no dynamical assumptions other than
the fact the current quark masses are sufficiently light,
and that expectation values of physical operators do not diverge,
one can
show that in such a medium there must exist at least one
pseudo-Goldstone mode with an energy that scales with the average
current quark mass as $\overline{m}^d$ where $d \ge \frac{2}{3}$.  This
is in sharp contrast with the case of the vacuum or isoscalar matter,
where the energy of the Goldstone modes goes as $\overline{m}^{1/2}$.

Moreover, if one makes highly plausible assumptions about the behavior of the
chemical potential for the system as one varies the quark masses, then
one can show even more peculiar behavior. In the vacuum one has pairs
of charged pseudo-Goldstone excitations ($\pi^+$--$\pi^-$ or $K^+$--$K^-$).
As one goes to a dense isospin-asymmetric medium (such as dense neutron
matter), however,  given these assumptions one can show that
only one member of this pair
survives as a pseudo-Goldstone mode.  In addition, the
excitation of the surviving pseudo-Goldstone boson
is proportional to the current quark mass $m_q$
itself, rather than to $\sqrt{m_q}$, as in the vacuum case.
Specifically, we
find that in the chiral limit in neutron matter there exist 1) a
pseudo-Goldstone mode
with quantum numbers $\pi^+$ with excitation energy~$\sim m_q$, 2)
pseudo-Goldstone mode
with quantum numbers $\pi^0$ with excitation energy~$\sim \sqrt{m_q}$, and
3) there are no pseudo-Goldstone modes with quantum number of $\pi^-$.

Our analysis is similar in many respects
to that of  Ref.~\cite{LSW:ref}, however
that work implicitly assumes that
the medium has low density. Mathematically, this equivalent to
expanding in density before expanding in the quark mass.
This is clearly appropriate
for mesonic atoms, where the meson feels only the tail of the nucleus
at a fraction of the nuclear saturation density, $\rho_0$.
Here we are interested in studying sufficiently dense systems so that
it becomes appropriate to
consider {\em chiral limit with the
density kept constant}. We will present some model estimates
indicating that this situation may be realized in nature if the density
of matter is of the order of a few $\rho_0$.
We will also demonstrate explicitly for a simple model that the chiral
and low (isovector)
density limits do not commute: an expansion in the small parameter
associated with one limit is singular in the expansion
parameter associated with
the other limit. Thus, physics associated with the two limits will be
quite different.

The paper is organized as follows: First, we derive our general result
of existence of unusually soft pseudo-Goldstone modes in nonsymmetric
nuclear medium. Next, we present a simple model illustrating that behavior
and give some numerical estimates of where we might see these
unusually soft modes. Finally, we discuss the case of kaons in
non-strange matter.

We find it useful to introduce the function
${\rm dim}_{\chi}(X) = \lim_{\overline{m} \to 0}
(\overline{m} \frac{d \log{X}}{d \overline{m}})$,
which we call the {\em chiral dimension} of the quantity $X$.
If near the chiral limit $X$ scales with the quark mass
as a power law in $\overline{m}$, {\em i.e.} if $X \sim
\overline{m}^\alpha$, then function
${\rm dim}_{\chi}(X)$ extracts the power $\alpha$.
For instance, in the vacuum ${\rm dim}_{\chi}(m_\pi)=1/2$ and
${\rm dim}_{\chi}(F_\pi)=0$.
The result is not modified by the presence of logarithms
%or poly-logarithms,
{\em i.e.}
${\rm dim}_{\chi}(\overline{m}^\alpha \log^n{\overline{m}})=\alpha$.

The starting point of our analysis is similar as in the
derivation of the standard GMOR relation. We examine  the expression for
the double commutator of the QCD hamiltonian density at point
$x=0$, ${\cal H}(0)$, with
the axial vector charges $Q_5^a = \int d^3 x J^a_{5,0}(x)$, where
$J^a_{5,\mu}$ is the axial vector current.
This expression can be evaluated
directly through the use of commutation rules for the quark fields:
\begin{equation}
\label{doublecomm}
\left [ Q_5^a, \left [ Q_5^a, {\cal H}(0) \right ] \right ] =
\overline{\psi}(0) \left \{ \lambda^a/2, \left \{ \lambda^a/2, M
\right \} \right \} \psi(0) \; , \;\;\;\;\; {\rm any~} \; a=1,...,8 \; ,
\end{equation}
where $\lambda^a$ are the Gell-Mann flavor matrices,
$M={\rm diag}(m_u, m_d, m_s)$ is the quark mass matrix, and
curly brackets denote anti-commutators.
Next, we take the matrix element of both sides of the above
operator relation in some physical state $|C \rangle$.   In all of  our
applications this state will  be taken to be
{\em spatially uniform}. The matrix
element of the RHS of Eq.~(\ref{doublecomm}) yields a linear
combination of quark condensates of various flavors.
For simplicity of notation, let us consider
the case of two flavors, with $m_u = m_d = \overline{m}$. The
generalization to 3 flavors is straightforward. We then
obtain
\begin{equation}
\label{matelement}
\langle C |
\left [ Q_5^a, \left [ Q_5^a, {\cal H}(0) \right ] \right ]
| C \rangle =
\overline{m} \langle C | \left ( \overline{u}u(0)+\overline{d}d(0)
\right ) | C \rangle
\equiv  \overline{m} \langle \overline{q}q \rangle_C \;, \;\;\;\;
a=1,2,3.
\end{equation}
On the other hand, the
LHS of the preceding equation may be evaluated by inserting a complete set
of intermediate states, yielding a sum rule.

Since the system $|C \rangle$
is spatially uniform, these intermediate states can
be labeled by their three-momentum, and some
additional label $j$. We denote them as $|j, \vec{p} \rangle$.
It is convenient to measure the three-momentum,
$\vec{p}$,
and the energy, $E_j$, of the intermediate states
{\em relative} to the ground state $|C \rangle$ at rest.
The total energy and the total three-momentum
of the state $|j, \vec{p} \rangle$ and of the
ground state $|C \rangle$ both form Lorentz four-vectors. Their difference
is therefore also a four-vector, and the unity can be
decomposed in the following Lorentz-invariant way:
\begin{equation}
\label{unity}
1=\sum_j \int \frac{d^3 p}{(2 \pi)^3\; 2 |E_j|}
|j,\vec{p} \rangle \langle j,\vec{p}|.
\end{equation}
Inserting Eq.~(\ref{unity}) inside the LHS of Eq.~(\ref{doublecomm}), and
using the fact that the ground state and the intermediate states
are eigenstates of the hamiltonian $H = \int d^3 x {\cal H}(x)$, namely
\mbox{$H|C \rangle = E_C|C \rangle$} and
$H|j, \vec{p} \rangle = (E_C+E_j)|j, \vec{p} \rangle$, we obtain
the following relation:
\begin{equation}
\label{relation1}
- \sum_j \frac{E_j}{|E_j|}
\left | \langle C|J^a_{5,0}(0)
|j, \vec{p}=0 \rangle \right |^2
= \overline{m} \langle \overline{q}q \rangle_C \; ,
\end{equation}
We may use the following operator relation
%
% *** check for factors of i ***
%
\begin{equation}
\label{PCACrel}
\partial^\mu J^a_{5,\mu}(0) = \left [ Q^a_5, {\cal H}(0) \right ]
\equiv \overline{m} D^a(0), \;\;\;\;\;\;
D^a(0) = \overline{\psi}(0) i \gamma_5 \tau^a \psi(0) \;\; ,
\end{equation}
to relate the matrix elements of $J^a_{5,0}$ to the matrix elements of $D^a$.
We find
%
% *** check for factors of i ***
%
\begin{equation}
\label{J2D}
\langle C|J^a_{5,0}(0)|j, \vec{p}=0 \rangle =
-i \frac{\overline{m}}{E_j}
\langle C|D^a(0)|j, \vec{p}=0 \rangle .
\end{equation}
Using Eq.~(\ref{J2D}) we may rewrite Eq.~(\ref{relation1}) in the
equivalent form which is familiar from the derivation of the GMOR
relation:
\begin{equation}
\label{relation2}
\sum_j \frac{\overline{m}}{|E_j| E_j}
\left | \langle C | D^a(0)
| j, \vec{p}=0 \rangle \right |^2
= - \langle \overline{q}q \rangle_C \; .
\end{equation}

Let us now recall the usual arguments leading from the preceding
general result to the GMOR relation.
In the GMOR case, the state $|C \rangle$ is
simply the vacuum.  First, we note that all excitation energies $E_j$ are
positive, since the vacuum is, be definition, the lowest-energy state
of the system. Therefore all
components of the LHS are positive.  Next, consider taking the limit
$\overline{m} \rightarrow 0$ on both sides of
Eq.~(\ref{relation2}). The RHS is assumed  to go to a nonzero constant
as a result of the spontaneous chiral symmetry breaking,
{\em i.e.} \mbox{${\rm dim}_{\chi}(\langle \overline{q} q \rangle) = 0$}.
On the LHS
only the terms with pseudo-Goldstones can contribute. This is
because
${\rm dim}_{\chi}(\langle 0| D^a(0) | j, \vec{p}=0 \rangle = 0$, and
thus only the states for which ${\rm dim}_{\chi}(E_j) = 1/2$ can contribute
to the sum rule in the chiral limit.
In the vacuum, $E_j$ is just the mass of the excited particle (pion).
Furthermore, from isospin symmetry
we have $m_{\pi^0} = m_{\pi^+} = m_{\pi^-}$.

In the present case our state $|C \rangle$ is not the vacuum.
Rather, we take  $| C \rangle$ to be the ground state of
the system subject to a {\em constraint} which requires that
the expectation value of some local (space-independent)
operator $c(x)$ (which commutes with $H$) has a fixed value, {\em i.e.}
$\langle C | c(x) | C \rangle = c = {\rm const}$.
In our case we are interested in medium with a given flavor
density, and $c(x) = \mu_u u^+u + \mu_d d^+d$ or
$c(x) = 3 \mu_B (u^+u + d^+d) + \mu_{I=1} (u^+u - d^+d)$, where
$\mu$'s are the corresponding chemical potentials for up and down
flavors, or alternatively for the baryon
number density and the isovector density
operator $u^+ u - d^+ d$.
If $c \neq 0$, then we have no
guarantee that $E_j$ is positive definite. The operator $J^a_{5,0}$ may
connect to states with a lower energy.

There are two distinct cases to consider:
1) the operator $c(x)$ commutes with the axial vector charges  or 2)
the  operator $c(x)$  does not commute
with the axial vector charges. As already
noted in Ref.~\cite{paperI}, if  $c(x)$
commutes with $Q_5^a$ (or equivalently with $D^a$), then the states
with negative $E_j$ do not contribute in the sum rule. This is simply
because in this case
$D^a$ connects $| C \rangle$ only to intermediate states with
the same value of $c$, and, by definition,
$| C \rangle$ is the {\em lowest} energy state of a given
value of $c$.

One physical situation where the constraint commutes with the $D^a$
operators is the
case of finite baryon density with zero isovector density (isospin symmetric
nuclear matter).  In this case the conclusions from the sum rule
(\ref{relation2}) are obtained analogously to the vacuum case, and,
so long as $\langle \overline{q} q \rangle_C \neq 0$
in the chiral limit, we
conclude there must exist pseudo-Goldstone  excitations
with $E_j \sim \sqrt{\overline{m}}$ with quantum numbers of
$\pi^0$, $\pi^+$ and $\pi^-$.\footnote{The case where
$\langle \overline{q} q \rangle_C = 0$ is interesting in its own right.
This problem is discussed in detain in ref. \cite{paperI}}
Isospin symmetry causes the neutral and
charged excitations to be degenerate in energy.
Thus, from the point of view of the GMOR relation, symmetric
nuclear matter behaves similarly to the vacuum.

Now we come to the main topic of our considerations.
Take nonsymmetric nuclear medium, {\em i.e.}
$\rho_{I=1} \equiv \langle C | (u^+u - d^+d) | C \rangle \neq 0$.
The neutral pseudo-Goldstone excitation still behaves in the usual way,
since the neutral axial vector charge commutes with the third isospin
component of the isospin charge, $Q_5^3$. Hence, as already remarked in
\cite{paperI}, even in nonsymmetric matter we have a
neutral pseudo-Goldstone excitation such that
$E_{\pi^0} \sim \sqrt{\overline{m}}$.
The case of charged pionic excitations, however,  is radically
different for two reasons.  The first is that
the isovector constraint does not commute with the charged
axial vector charges $Q^1_5$ and $Q^2_5$.  The second difference is
related to the existence of a
second sum rule which is trivially zero for an isoscalar medium but not
for a medium with nonzero isospin density.
This sum rule is derived rather easily.
Using the fact that
$\left [ Q^a_5, J^b_{5,0}(0) \right ] = i \epsilon^{abc} J^c_0(0) $, where
$J^a_\mu$ is the vector current, and inserting Eq.~(\ref{unity}) inside
the commutator, we obtain
\begin{equation}
\label{rho}
\rho_{I=1} = \sum_{j_-} \frac{\overline{m}^2}{| E_{j_-} | E_{j_-}^2}
| \langle j_-, \vec{p} | D^-(0) | C \rangle |^2 -
 \sum_{j_+} \frac{\overline{m}^2}{| E_{j_+} | E_{j_+}^2}
| \langle j_+, \vec{p} | D^+(0) | C \rangle |^2 \;\; ,
\end{equation}
where we have decomposed the sum over $j$ into two classes of states:
those with isospin
one unit more ($j_+$) or less  ($j_-$) than in $| C \rangle$.
Equation (\ref{relation2}) may be decomposed similarly, giving
\begin{equation}
\label{qbarq}
- \langle \overline{q}q
\rangle_C
= \sum_{j_-} \frac{\overline{m}}{2 | E_{j_-} | E_{j_-}}
| \langle j_-, \vec{p} | D^-(0) | C \rangle |^2 +
\sum_{j_+} \frac{\overline{m}}{2 | E_{j_+} | E_{j_+}}
| \langle j_+, \vec{p} | D^+(0) | C \rangle |^2 \;\;  .
\end{equation}

Now consider the sum rule given in Eq.~(\ref{rho}) as one
approaches the chiral limit.  In isovector matter, the left hand side
of this sum rule is nonzero, by definition.  Accordingly, at least one
term in this sum rule must be nonzero.  However, all terms in the
sum rule go as $ \overline{m} /E_j^3 $ times matrix elements.
Since the matrix elements are assumed to be finite in the chiral limit,
one sees that the
only way any term can make a  contribution in the chiral limit is
if ${\rm dim}_{\chi}(E_j) \ge 2/3$.
This completes the demonstration of the first
point of this Letter. Note, that this relation is written as
inequality rather than an equality since it is possible that the
chiral dimension of the matrix element could well be $ >0$.
Indeed, as we will show below, given reasonable assumptions it is
in fact $ >0$.

To proceed further we need to make the assumption that
the chiral dimension of the isovector
chemical potential ${\rm dim}_{\chi}(\mu_{I=1})=0$.
Since we are implicitly
taking the chiral limit at a fixed isospin
density $\rho_{I=1}$, we can treat $\rho_{I=1}$ as an
{\em external parameter}, independent of the chiral parameter --- the
assumption which we must make is that the difference in energy density
between this state and the
the lowest isospin symmetric state is independent of $\overline{m}$.
Although we cannot prove that ${\rm dim}_{\chi}(\mu_{I=1})=0$ from
first principles, we can present the following
physical argument in its favor:
The isovector interaction has a contribution
produced by the $\rho$-meson exchange. Suppose we place an object of
isospin $I_3$ in the isovector medium. The interaction is
$(g_\rho^2/m_\rho^2)\;\rho_{I=1}I_3$, and the corresponding
chemical potential is $(g_\rho^2/m_\rho^2)\;\rho_{I=1}$. Here
$g_\rho$ and $m_\rho$ are $\rho$-meson coupling constant and mass
in the medium. Their chiral dimension in the vacuum is $0$, and,
unless something very unusual happens, this is also true
in medium. Therefore $\rho$-exchange produces
${\rm dim}_{\chi}(\mu_{I=1})=0$. It is very unlikely that this result
could be altered by other processes --- they would have to exactly cancel
the $\rho$-exchange mechanism.

Equipped with the assumption that ${\rm dim}_{\chi}(\mu_{I=1})=0$, we can
continue the analysis of the sum rules (\ref{rho}) and (\ref{qbarq}).
Suppose for definiteness that $\mu_{I=1} > 0$ (as it is for neutron matter).
{}From definition of the chemical potential as the minimum
energy needed to lower the isospin by one unit, we have
\begin{equation}
\label{mu}
E_{j_\mp} = \mu_{I=1} \pm \delta E_{j_\mp} \;\;\; ,
\;\;\;\;\; {\rm with~} \; \delta E_{j_\mp} \ge 0 \; .
\end{equation}
Therefore the energies of states with one unit of isospin less
than $|C \rangle$, $E_{j_-}$, can never go to zero. In the
chiral limit only the states with $E_j \to 0$ can contribute
to the sum rules (\ref{rho}) and (\ref{qbarq}), and as a result only
the states $j_+$ contribute.   Note, this is very different from the
vacuum case --- it shows that in neutron matter the $\pi^+$
modes can be pseudo-Goldstone modes while the $\pi^-$
cannot.
Restricting the sum rules to the $j^+$ modes only gives
\begin{equation}
\label{chirsrI}
\rho_{I=1} = \lim_{\overline{m} \to 0 } \left (
- \sum_{j_+} \frac{\overline{m}^2}{| E_{j_+} |^3}
| \langle j_+, \vec{p} | D^+(0) | C \rangle |^2 \right ) \;\; ,
\end{equation}
and
\begin{equation}
\label{chirsrII}
 \lim_{\overline{m} \to 0 }
\left (\langle \overline{q}q
\rangle_C \right )
=  - \lim_{\overline{m} \to 0 }
\left (
\sum_{j_+} \frac{\overline{m}}{2 | E_{j_+} | E_{j_+}}
| \langle j_+, \vec{p} | D^+(0) | C \rangle |^2 \right )\;\; .
\end{equation}
Now, the sum rule (\ref{chirsrI}) contains only semi-negative
contributions. Comparing the chiral dimensions on both sides,
it follows that there must exist a mode for which
\begin{equation}
\label{dimI}
0 = 2 - 3\; {\rm dim}_\chi (E_{j_+}) + 2\; {\rm dim}_\chi
(\langle j_+,\vec{p}=0 | D^+(0) | C \rangle) \;.
\end{equation}
The sum rule (\ref{chirsrII}) may contain both positive and negative
contributions, since the sign of $E_{j_+}$ is not restricted.
That means that in principle there may be cancellations of the leading
chiral dimension on the RHS of Eq.~(\ref{chirsrII}), and we
obtain the following inequality:
\begin{equation}
\label{dimII}
{\rm dim}_\chi (\overline{q}q) \ge
1 - 2\; {\rm dim}_\chi (E_{j_+}) + 2\; {\rm dim}_\chi
(\langle j_+,\vec{p}=0 | D^+(0) | C \rangle) \;.
\end{equation}
The conjunction of conditions (\ref{dimI}) and (\ref{dimII}) gives
${\rm dim}_\chi (E_{j_+}) \le
1 + {\rm dim}_\chi (\langle \overline{q}q \rangle)_C$.
The inequality becomes equality unless there is an exactly cancellation
of the leading order contribution
on the RHS of Eq.~(\ref{chirsrII}). For instance,
the equality is automatically the case if there is only one state
$|j_+,\vec{p}=0 \rangle$
which becomes a pseudo-Goldstone mode.  Also, potential cancellations
are not associated with any symmetry, hence it is difficult
to imagine that the leading chiral dimension can indeed be exact
canceled  on the RHS of Eq.~(\ref{chirsrII}). With no cancellations we have
\begin{equation}
\label{eq}
{\rm dim}_\chi (E_{j_+}) =
1 + {\rm dim}_\chi (\langle \overline{q}q \rangle_C) \; .
\end{equation}
If we are in the spontaneously broken phase, then we expect
that ${\rm dim}(\langle \overline{q}q \rangle_C) = 0$, as in the vacuum.
Clearly, this dimension has at least to be positive semi-definite, if
the chiral condensate is to be well-behaved in the chiral limit.
This leads to the following final result:
\begin{equation}
\label{eq2}
{\rm dim}_\chi (E_{j_+}) =  1 \;, \;\;\;\; {\rm dim}_\chi (D^+) = 1/2 \; ,
\;\;\;\; {\rm dim}_\chi (J^+_{5,0}) = 1/2 \; ,
\end{equation}
where the third equality follows from Eq.~(\ref{J2D}).
Note that this behavior is radically different than in the vacuum,
where we have
\begin{equation}
\label{eqvac}
{\rm dim}_\chi (m_\pi) =  1/2 \;, \;\;\;\; {\rm dim}_\chi (D^+) = 0 \; ,
\;\;\;\; {\rm dim}_\chi (J^+_{5,0}) = 1/2 \; .
\end{equation}
Of the three pionic  pseudo-Goldstone modes in the vacuum, only
only 2 survive as pseudo-Goldstone modes in dense isovector matter. For
negative $\rho_{I=1}$
the negative charge excitation disappeared,
the positive charge
excitation became unusually soft, $E_{\pi^+} \sim \overline{m}$, and the
neutral
excitation retained its chiral dimension,
$E_{\pi^0} \sim \sqrt{\overline{m}}$. If $\rho_{I=1} > 0$, then
the positive and negative excitation change the roles.

The fact that only one member of the pair of
charged
pseudo-Goldstone mesons
remains a pseudo-Goldstone mode in dense isospin-asymmetric matter
is quite natural. The medium
itself breaks the symmetry between the members of the pair --- thus one
would be quite surprised if they had the same energy.   One expects the
symmetry breaking induced the medium to split the degenercy between the
two states.  However, as one approaches the chiral limit a
pseudo-Goldstone mode must go to zero excitation energy.  Thus, if both
modes were pseudo-Goldstone modes they
would both have to approach zero and hence would
become degenerate.

Before giving an illustrative example, let us
recapitulate our derivation, listing all assumptions made on the way.
First, let us stress that what we were after is a fundamental result
derived without any explicit reference to dynamics, microscopic
structure of the modes, {\em etc}. In matter these modes are undoubtedly
quite complicated, involving particle-hole excitations, {\em etc}.
We have not made any dynamical assumptions.
These were our key ingredients:

\begin{enumerate}

\item The chiral limit is taken first, while the isovector density is
kept constant.

\item ``Reasonableness'' conditions are that expectation values
of operators, {\em e.g.} $\langle \overline{q}q \rangle_C$, do not
diverge in the chiral limit. If this were not true, than
the chiral limit would not make sense in the isovector matter.

\item The isovector chemical potential is assumed to be nonzero
in the chiral limit.

\item We also assumed that there are no exact cancellations of the leading
chiral powers
in the sum rule
(\ref{qbarq}). This assumptions is equivalent to having no (accidental)
cancellations from a priori possible multiple branches of pseudo-Goldstone
modes. If there is only one such branch, this result follows trivially.

\end{enumerate}

\noindent
Assumptions 1) and 2) are sufficient to show that
there exists a pseudo-Goldstone mode for
which ${\rm dim}_\chi (E_j) \ge 2/3$.
Assumptions 1) -- 4) give ${\rm dim}_\chi (E_j) = 1$ for this mode.

Now, let us present a simple model which will illustrate the behavior
of pseudo-Goldstone modes in nonsymmetric medium.
Consider the pion propagating in an isovector medium and interacting
with it through the $\rho$-meson exchange.   Moreover, in this toy
model,  we will assume that the $\rho$-meson exchange is the {\em only}
interaction between the pions and the medium.  As is well known\cite{Bhaduri},
if all vector-isovector interactions are mediated by the $\rho$ meson,
then consistency of the $\rho$ exchange picture with the various
soft-pion theorems requires the KSFR \cite{KSFR} relation
$m_\rho^2/g_\rho^2 = 2 F_\pi^2$ to be exact.
The inverse propagator for the $\pi^\pm$ mesons with four-momentum $p$ is
$(G(p)^\pm)^{-1} = p^2 - m_\pi^2 \mp p_0 A$, where
$A = (g_\rho^2 / m_\rho^2) \rho_{I=1}$. The (positive energy) poles of
$G(p)$ for the pion at rest ($\vec{p}=0$) occur at
\begin{eqnarray}
E^+ & =  & - A/2 + \sqrt{A^2/4 + m_\pi^2} \nonumber\\
E^- & =  & A/2 + \sqrt{A^2/4 + m_\pi^2}  \label{modes}
\end{eqnarray}
In the chiral limit, $E^+ \to m_\pi^2/A$ --- this is the unusually soft
pseudo-Goldstone mode, with $\dim(E^+) = 1$, and $E^- \to A$ --- this
is no longer a pseudo-Goldstone mode, since $\dim(E^-) = 0$.
This is precisely the behavior which we predicted.
Obviously, the neutral pion
in unchanged by the isovector interactions, and
$\dim_\chi (E^+) = 1/2$.

Let us also examine the matrix elements of the operator $D^\pm$.
We have, from definition (\ref{PCACrel}),
$D^\pm = \partial^{\mu} J^\pm_{5,\mu} / \overline{m} =
(F_\pi m_\pi^2 / \overline{m})\; \pi^\pm$,
where $\pi^\pm$ is the (charged) pion interpolating field, and
$F_\pi$ is the pion decay constant in the vacuum.
Let abbreviate the states corresponding to excitations $E^\pm$ by
$|\pm \rangle$. Using
$|\langle \pm | \pi^\pm | C \rangle |^2 =
\lim_{p_0 \to E^j} 2 E^j (p_0 - E^j) G^\pm(p_0,\vec{p}=0)$, where
$j=+$~or~$-$,
we find that in the chiral limit
$| \langle +| D^+ | C \rangle |^2 \to 2 F_\pi^2 m_\pi^6 /
(A^2 \overline{m}^2) \sim \overline{m}$, in compliance to the
general result (\ref{dimI}), and
$| \langle - | D^+ | C \rangle |^2 \sim 1$.
Only the positive charge mode saturates the sum rules
(\ref{rho}) and (\ref{qbarq}). Explicitly, we get
\begin{equation}
\label{modelI}
\rho_{I=1} = F_\pi^2 A = 2
\left ( \frac{g_\rho F_\pi}{m_\rho} \right )^2 \rho_{I=1} \; ,
\end{equation}
and
\begin{equation}
\label{modelII}
\lim_{\overline{m} \to 0} \langle \overline{q}q \rangle_C =
- \frac{F_\pi^2 m_\pi^2}{\overline{m}} \;.
\end{equation}
Equation (\ref{modelI}) means that, as expected, consistency in the
chiral limit requires the KSFR relation to be exact.
Relation (\ref{modelII}) also shows the formal consistency, since
exactly the same
equation is obtained by considering the sum rule for the neutral pion.  Since
the neutral pion is not affected by the $\rho$-exchange, we immediately get
Eq.~(\ref{modelII}).

Under what circumstances is the analysis here useful? --- {\it i.e.} at
what isospin density does the
system go from effectively having three
pseudo-Goldstone modes whose energies all have chiral dimension 1/2 to
having two pseudo-Goldstone modes whose energies have
chiral dimension of 1 and 1/2.  Formally, the analysis
is valid for any nonzero isospin density {\em provided we go to the
$m_q \rightarrow 0$ limit}.
On the other hand, in nuclear physics the value of
$m_q$, though small, is nonzero.  Thus, the question
of interest is under what circumstances is the quark
mass small enough so that  the results of $m_q \rightarrow 0$ limit are
close to the physical results.
The toy model considered above, although probably quite
unrealistic, gives us considerable insight.

Consider Eq.~(\ref{modes}); one approaches the
limit in which the present analysis applies when
$ |A|/2 \gg m_{\pi}$.  We
note that in this toy model $A=(\rho_{I=1} g_\rho^2) / m_\rho^2$
is the chemical potential associated
with isospin, $\mu_{I=1}$.
Also in this simple model, $m_{\pi}$ is the energy of the
pseudo-Goldstone mode when $ \rho_{I=1}=0$.
In the general case it may be different from $m_\pi$, so let us
denote it by $E_0$.  Thus, a reasonable
general criterion for the region of applicability of our
analysis is when
$\mu_{I=1} \gg E_0$.
In our toy model, this condition is equivalent to
$|\rho_{I=1}| \gg \overline{\rho}
\equiv 4 F_{\pi}^2 m_{\pi} \simeq 3.7 \rho_0$, where
$\rho_0$ is the nuclear saturation density. The above estimate used
the vacuum values of $F_\pi$ and $m_\pi$. However, this probably
leads to an overestimation of $\overline{\rho}$, because
matter with large isovector density also has large baryon
density, $\rho_B$, which considerably reduces the value of $F_\pi$ from
its vacuum value
\cite{DL:ref,CFG:ref,LKW:ref,CSDZ:ref,ME:ref,CM:ref,BM:ref,MB:ref}.
In turn, $\overline{\rho}$ should be
substantially reduced by the presence of $\rho_B$.

We also note one more fact illustrated by our toy model.
It is apparent that the chiral limit and the low-isovector-density
limit do not commute.
In our model in the chiral limit the expansion  parameter is
$\alpha = m_\pi^2 F_\pi/\rho_{I=1}$
--- this is singular in the isovector density.
Conversely, in in the low density limit the expansion parameter
is $1/\alpha$, and this, it turn, is singular in $m_\pi$.
We believe this is a manifestation of a general result.

Now let is turn our attention to kaons, since recently the possibility
of $S$-wave kaon condensation has been extensively discussed
\cite{MPW,PW,KN,BKRT,BLRT,YMK:ref,MB:ref}.
The meaning of the chiral limit in this case is somewhat subtle and
this subtlety greatly affects the applicability of our analysis.
As the
current quark masses tend to zero (including $m_s$) two
phenomena happen in the nuclear medium. One is the change of properties of
particles, which is the subject of our analysis.
The second phenomenon is the change of the ground state of the
medium itself.

In the strict
$SU(3)$ chiral limit the ground state of matter has equal amount of up,
down and strange quarks.  In that case the octet of axial vector
charges commutes with the (baryon number) constraint, and pions,
kaons and $\eta$ are the usual pseudo-Goldstone excitations with
$E^a \sim \sqrt{m_q}$. This is not the situation of
relevance
at moderately low densities.   At first sight, one might rather keep the matter
non-strange, by imposing the constraint
$\mu_u u^+u + \mu_d d^+d$, and watch the formal behavior of the kaon
as $m_s \to 0$. In this case the $a=4,...,7$ components of the axial
vector charge do not commute with the constraint.
The situation is
analogous to the case of the charged pion in nonsymmetric medium. We
only have to change the usual I-spin into U-spin or V-spin.
Again, only two kaon-like excitation (out of 4 in the vacuum)
are pseudo-Goldstone modes,
with excitation energy $\sim m_q$.

The preceding argument, however, is formal; it  depends on using
$\mu_u u^+u + \mu_d d^+d$ as  the constraint.  This is not the relevant
constraint for the neutron matter calculations---at least in the
interesting  applications to neutron stars.  In that case of interest
the constraints are on the net baryon density and the net charge
density \cite{MPW,PW,KN,BKRT,BLRT}.  The reason that these are
the appropriate constraints is clear---in the neutron star case
one is working at long times scales for which the system can be
expected to be in equilibrium with respect to weak interaction
processes which can change the flavor quantum numbers.  Accordingly,
the analysis given above does not give any quantitative insight into the
problem of kaon condensation.

In conclusion, we have shown the behavior of pseudo-Goldstone modes in
dense isospin-asymmetric nuclear matter is very different from the case
of the vacuum or isospin symmetric matter.  Of
the three pionic modes which exist in the vacuum only  two remain in
dense isovector matter.  Moreover, of these remaining modes, the charged
mode is anomalously light---having a chiral dimension of one rather than
one half.

This work has been supported by the NSF--Polish Academy of
Science grant \#INT-9313988, NSF PYI grant \#PHY-9058487, DOE grant
\#DE-FG02-40762, and by the Maria Sk\l{}odowska-Curie grant
\#PAA/NSF-94-158.


\begin{references}

\bibitem{mesons:in:medium} See for example T. E. O. Ericson and W. Weise,
{\em Pions and Nuclei}, Oxford Science Pub. (1988).

\bibitem{GMOR:ref}  M. Gell-Mann, R. Oakes and B. Renner,
Phys. Rev. {\bf 175} (1968) 2195.

\bibitem{LSW:ref} M. Lutz, A.Steiner and W. Weise,
Nucl. Phys. {\bf A 574} (1994) 755.

\bibitem{paperI} T. D. Cohen and W. Broniowski, U. of Maryland
preprint DOE-ER-40762-043,hep-ph/9407354, to appear in Phys. Lett. B.

\bibitem{Bhaduri} See for example R. K. Bhaduri,
{\em Models of the Nucleon, from Quarks to Soliton},
Addison-Wesley (1988), Lecture Notes and Supplements in Physics.

\bibitem{KSFR} K. Kawarabayashi and M. Suzuki, Phys. Rev. Lett.
{\bf 16} (1966) 255; Riazuddin and Fayazuddin, Phys. Rev. {\bf 147}
(1966) 1071.

\bibitem{DL:ref} E. G. Drukarev and E. H. Levin, Nucl. Phys.
{\bf A 511} (1990) 679; {\bf A516} (1990) 715(E);
Prog. Part. Nuc. Phys. {\bf 27} (1991) 77.

\bibitem {CFG:ref} T. D.  Cohen, R. J. Furnstahl, D. K. Griegel,
Phys. Rev.  C  {\bf 45} (1992) 1881; Phys. Rev. Lett. {\bf 67} (1991) 961.

\bibitem{LKW:ref} M. Lutz, S. Klimt, W. Weise,
Nucl. Phys. {\bf A 542} (1992) 521.

\bibitem {CSDZ:ref} L. S. Celenza, C. M. Shakin,
W. Dong and X. Zhu, Phys. Rev. C {\bf 48} (1993) 159.

\bibitem{ME:ref} M. Ericson, Phys. Lett.  {\bf B301}, 11 (1993).

\bibitem{CM:ref} G. Chanfrey and M. Ericson, Nucl. Phys. {A556}, 427 (1993).

\bibitem{BM:ref} M. C. Birse and J. A. McGovern, Phys. Lett
{\bf B309}, 234 (1993).

\bibitem{MB:ref} M. C. Birse, University of Manchester preprint MC/TH 94/13,
nucl-th/9406029.

\bibitem{MPW} D. Montano, H. D.  Politzer and M. B. Wise, Nucl. Phys.
{\bf B375} (1992) 507.

\bibitem{PW} H. D.  Politzer and M. B. Wise,
Phys. Lett. {\bf B273} (1991) 156.

\bibitem{KN} D. B. Kaplan and A. E. Nelson,
Phys. Lett. {\bf B175} (1986) 57; {\bf B192} (1987) 193.

\bibitem{BKRT} G. E. Brown, K. Kubodera, M. Rho,
V. Thorsson, Phys. Lett. {\bf B291} (1992) 355.

\bibitem{BLRT} G. E. Brown, C.-H. Lee, M. Rho,
V. Thorsson, Nucl. Phys. {\bf A567} (1994) 937.

\bibitem{YMK:ref} H. Yabu, F. Myhrer, K. Kubodera,
Phys. Rev. {\bf D 50} (1994) 3549.


\end{references}
\end{document}